\documentclass[aps,prl,twocolumn,groupedaddress,showpacs]{revtex4}
\usepackage{graphicx}
\graphicspath{{figure file fold path}}
\usepackage[dvipsnames,usenames]{color}
\usepackage{color}
\usepackage{amsmath}
\usepackage{float}
\usepackage{amssymb}
\usepackage{amsthm}
\usepackage{mathrsfs}
\usepackage{booktabs}  
\usepackage{ragged2e} 
\usepackage{tabularx}
\usepackage{multirow}

\emergencystretch=\maxdimen
\begin{document}

\title{Polarization Lifetime as a Design Axis for Task-Adaptive Photonic Reservoir Computing}
\author{Tao Wang$^{1}$, and Kathy L\"udge$^{2}$}

\affiliation{$^1$State Key Laboratory of Integrated Service Networks, School of Telecommunications Engineering, Xidian University, Xi'an 710071, China}
\affiliation{$^2$Technische Universit\"at Ilmenau, Institute of Physics, 98693 Ilmenau, Germany}

\date{\today}

\begin{abstract}
Photonic time-delayed reservoir computing (TDRC) systems based on conventional Class B lasers are intrinsically limited by fragile external cavity mode (ECM) locking dynamics. Here, we demonstrate that engineering the polarization lifetime \((T_2)\) alongside the carrier lifetime \((T)\) fundamentally reshapes the computational topology and stability of TDRC. Through a systematic comparative study of four laser classes (Class A, Class B, Class \(C_A\), and Class \(C_B\)), we reveal that a finite polarization lifetime acts as an intrinsic dynamic filter, dissolving discrete ECM chaos grids into continuous, phase-tolerant operational basins. Information processing capacity (IPC) decomposition further exposes distinct computational profiles: the perfectly matched Class \(C_A\) achieves the highest nonlinear capacity (16.28) and the highest total capacity (19.29), whereas Class A exhibits the highest linear capacity (4.70). We identify a task-dependent reversal that governs hardware design: Class A and Class \(C_A\) achieve comparable, and the lowest, NRMSE on short-horizon Mackey--Glass prediction (0.0103 and 0.0105, within one standard deviation), while the time-scale-matched Class \(C_A\) dominates the higher-dimensional Lorenz (0.0506) and the experimental Santa Fe (0.4251) tasks. The ranking of the four classes by linear IPC closely matches the ranking by Mackey--Glass accuracy, and the ranking by nonlinear IPC matches the Lorenz ranking exactly and the Santa Fe ranking to within the mask-to-mask uncertainty. Under realistic quantum noise, the noise resilience is primarily dictated by the carrier lifetime \(T\) rather than the polarization lifetime \(T_2\): Class A retains the lowest NRMSE (0.47) even at the extreme limit of the spontaneous emission factor \(\beta = 10^{-1}\), whereas the long-carrier Class B suffers a systematic degradation (0.98). A residual \(T_2\)-mediated benefit of approximately \(20\%\) persists in the slow-carrier Class \(C_B\) (0.78 vs. 0.98). Our findings establish time-scale engineering as a nuanced design paradigm for next-generation optical computing, in which the optimal hardware configuration is determined not by a universal figure of merit but by the intrinsic time-scale hierarchy demanded by the target task.
\end{abstract}


\maketitle 

\section{Introduction}
The exponential growth of artificial intelligence, particularly in time-series forecasting and real-time control, has created an urgent demand for neuromorphic hardware capable of processing temporal data streams with ultrahigh throughput and minimal energy consumption~\cite{ahmed2025universal, shastri2021photonics}. Photonic reservoir computing (RC) has emerged as a promising paradigm to meet these demands~\cite{yan2024emerging}. By harnessing the transient, high-bandwidth dynamics of optical systems to project input signals into a high-dimensional state space, photonic RC circumvents the computationally expensive iterative training of recurrent neural network weights~\cite{appeltant2011information, tanaka2019recent}. In the widely adopted time-delayed RC (TDRC) architecture, the spatial complexity of a conventional network is elegantly replaced by the temporal multiplexing of a single physical nonlinear node subject to delayed optical feedback~\cite{huang2025stochastic, hasegawa2023parallel, Rohm2020}. Consequently, the computational capacity, memory depth, and noise resilience of the entire system are dictated entirely by the intrinsic physical dynamics of this single node.

To date, the vast majority of photonic TDRC implementations have relied on semiconductor lasers as the nonlinear node~\cite{Estebanez2023, wang2023reservoir, kanno2022reservoir, wang2024reservoir, appeltant2011information, brunner2013parallel, dong2024pssr}. Experimentally, photonic TDRC has achieved high-throughput processing at multi-gigabit rates using transients in delayed feedback systems~\cite{brunner2013parallel}. Theoretically, the rich dynamics of semiconductor lasers subject to optical feedback have been extensively modeled using delay differential equations to design optimal reservoir states~\cite{dong2024pssr}. The dynamics of such lasers are universally modeled by Class B rate equations, a simplified mathematical framework that assumes the macroscopic medium polarization decays infinitely fast compared to the cavity photon lifetime and the carrier inversion lifetime \(T\)~\cite{tredicce1985instabilities, roos2021stabilizing}. By adiabatically eliminating the polarization dynamics, Class B models reduce the system to a two-dimensional phase space dominated by relaxation oscillations~\cite{wieczorek2005dynamical, ludge2020laser}. While these lasers have been successfully applied to benchmark tasks, their pronounced, carrier-induced relaxation oscillations (which are typically underdamped in conventional semiconductor quantum-well lasers) impose fundamental constraints on the accessible dynamical responses. Most critically, the high-dimensional dynamics close to the edge of stability in Class B lasers are accessible only within narrow windows of a highly structured parameter space, where stable and chaotic dynamics alternate frequently owing to the complex locking of external cavity modes (ECMs). This operational intricacy implies that even minute thermal drifts or fabrication imperfections can push the system out of its optimal computational window, leading to a catastrophic collapse in performance. These fragilities stem from an inherent mismatch between the ultra-fast polarization and slow carrier relaxation timescales.

A promising route is to systematically engineer the laser's intrinsic timescales, deliberately retaining a finite polarization lifetime (T$_2$) that acts as a dynamical filter and a topological smoother. The adiabatic elimination of the polarization lifetime in Class B models completely discards a rich dynamical resource that could fundamentally alter this picture. This omission has been shown to be particularly significant in the context of the modulation response of nanolasers, where standard rate-equation approaches miss essential dynamical features~\cite{aust2016modulation}. When the polarization lifetime ($T_2$) becomes comparable to or longer than the cavity photon lifetime, the laser enters the coherent Class C regime, where polarization evolves as an independent dynamic variable~\cite{wieczorek2005dynamical}. Moreover, by decoupling $T_2$ from the carrier lifetime $T$, one obtains a distinct Class C$_B$ regime, where a long $T_2$ acts as a natural low-pass filter, imposing a smoothing effect on the complex dynamics of conventional Class B lasers. The role of such slow polarization dynamics as an intrinsic spectral filter has been previously discussed in the context of delay-coupled laser systems~\cite{luedge2019slowpolarization}. This raises a pressing and previously unanswered question: can the deliberate reintroduction of macroscopic polarization and its interplay with carrier lifetimes fundamentally transform the computational landscape of photonic TDRC? Specifically, can a finite polarization lifetime dissolve the rigid phase-matching constraints that render Class B reservoirs fragile, endow the system with intrinsic adaptability to diverse task timescales, and generate high-dimensional nonlinear mappings that are genuinely robust to spontaneous emission noise?

In this work, we answer this by conducting a systematic comparative study of four representative laser classes: Class A (fast \(T_2\), fast \(T\)), Class B (fast \(T_2\), slow \(T\)), Class C\(_A\) (finite \(T_2\), finite \(T\)), and Class C\(_B\) (finite \(T_2\), slow \(T\)). Here, it is necessary to mention that these labels do not strictly follow the classical Arecchi classification. As illustrated in Fig.~\ref{Figure_11}, our central thesis is that the deliberate retention of finite \(T_2\) acts as a geometric timescale toolkit for shaping the reservoir's state space. We perform comprehensive sweeps across feedback strength \(\kappa\), detuning \(\delta\omega\), and phase \(\phi\) to map the operational phase diagrams for each class, revealing that Class C$_B$ lasers benefit from polarization-induced filtering that significantly broadens their stable regions compared to the fragile locking grids of Class B, namely the strictly phase-matched stable continuous-wave (CW) operation locked onto external cavity modes (ECMs)~\cite{luedge2019slowpolarization}. Harnessing these findings, we lock each class to its optimal parameters, validate task-specific performance on Mackey-Glass (MG), and decompose the computational mechanisms via the rigorous Information Processing Capacity (IPC) framework. We also quantify noise resilience by scanning the spontaneous emission factor \(\beta\), demonstrating that while Class B's long carrier lifetime makes it vulnerable to noise accumulation, the natural filtering of Class C\(_B\) provides intrinsic robustness against quantum fluctuations. Collectively, our results establish that engineering laser timescales, particularly the retention of finite \(T_2\), is not merely a theoretical exercise, but a powerful practical lever for designing high-performance, noise-resilient, and task-adaptive photonic TDRC hardware.

\section{Theoretical Framework and Laser Dynamics}
The foundation of modern photonic RC rests on the rich and complex dynamics of laser systems, which serve as the physical reservoir for information processing. As illustrated in the overview diagram of Fig.~\ref{Figure_11}, semiclassical laser models can be broadly categorized into four fundamental operational regimes: Class A, Class B, Class C$_A$, and Class C$_B$~\cite{arecchi1984deterministic}. In addition, quantum dot (QD) lasers often exhibit dynamics that evolve in between Class A and Class B, characterized by relatively faster polarization relaxation than that in conventional quantum-well devices while retaining significant carrier-induced memory effects~\cite{arecchi1984deterministic}. The crucial distinction among these configurations lies in the deliberate retention or elimination of the macroscopic polarization lifetime \(T_2\), which fundamentally shapes the interplay between the field and the gain medium. A detailed understanding of these dynamics is essential for designing optimal reservoir computing architectures.

\subsection{The Unified Maxwell-Bloch Framework}
The dynamics of the laser-based reservoir are governed by the Maxwell-Bloch delay-differential equations (DDEs), which describe the self-consistent interaction of the complex optical field \(E(t)\), macroscopic polarization \(P(t)\), and carrier inversion \(N(t)\). The equations are expressed as:

\begin{eqnarray}
\frac{dE}{dt} &=& c P - \frac12 E + K E(t-\tau)e^{-i\phi} + \sqrt{\frac{\beta}{2}}\,\xi_E(t), \\
\frac{dP}{dt} &=& \frac{1}{T_2}\bigl[-(1+i\delta\omega)P + E N\bigr], \\
\frac{dN}{dt} &=& \frac{1}{T}\bigl[P_{\text{pump}}(t) - N - 2c\,\text{Re}(E^*P)\bigr],
\end{eqnarray}

\noindent Here \(E\) denotes the complex optical field. The coefficient $c$ represents the normalized cavity decay rate given by

\begin{eqnarray}
c = (\frac{2\delta\omega}{T_2+2})^2+1
\end{eqnarray}

\noindent which ensures the first laser threshold at \(P_{\text{pump}} = 0.5\) (\(\delta\omega\) is the frequency detuning between the cavity mode and the gain resonance). The term \(K E(t-\tau)e^{-i\phi}\) is the time-delayed feedback, which is essential for creating memory and coupling between virtual nodes. \(P\) is the macroscopic polarization with lifetime \(T_2\). This parameter fundamentally differentiates the laser classes: in Class A, \(T_2\) is nearly instantaneous (\(T_2 \ll T\)), effectively eliminating polarization dynamics, whereas in Class C\(_A\) and Class C\(_B\), a finite \(T_2\) (comparable to the cavity photon lifetime, such as those governed by spin-relaxation rates in semiconductor VCSELs or the prolonged photon lifetimes in high-\(\beta\) microcavities~\cite{deng2021physics}) preserves polarization as an independent dynamical variable. \(N\) is the carrier inversion with lifetime \(T\), which is driven by the modulated pump \(P_{\mathrm{pump}}(t) = P_{\mathrm{base}} \bigl(1 + \gamma\,u(t)M_N\bigr)\), where \(u(t)\) is the input signal, \(M_N\) is the periodic mask drawn from a uniform distribution on $[-1, 1]$, and \(\gamma\) controls the modulation depth. \(\xi_E(t)\) is a delta-correlated complex Gaussian noise term with strength \(\beta\)~\cite{wang2023reservoir}, describing spontaneous emission, $\beta = \beta_{noise}$.

\subsection{Time-Delayed Reservoir Computing (TDRC) Implementation}
\begin{figure*}[ht!]
\centering
  \includegraphics[width=14cm, height=7.0cm]{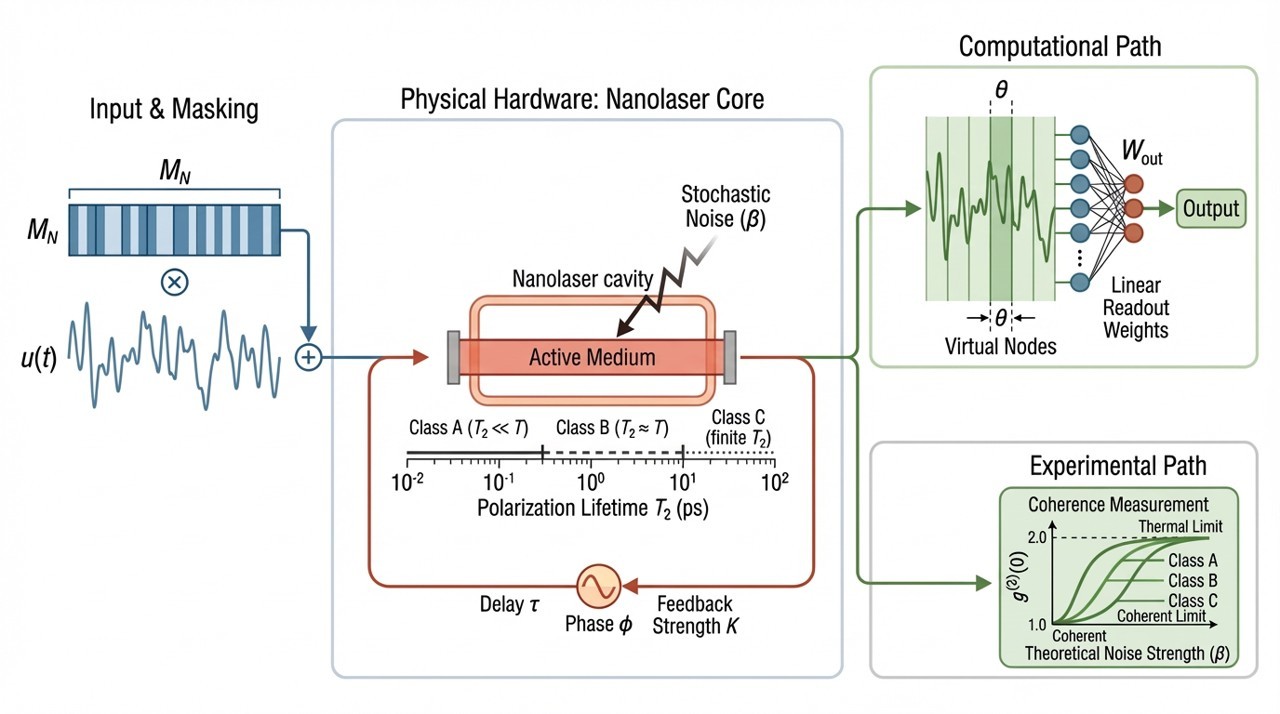}
  \caption{Architecture of the photonic TDRC system. The physical hardware core is a laser subject to delayed optical feedback. The four operational regimes - Class A, Class B, Class C$_A$, and Class C$_B$ - are defined by their intrinsic polarization ($T_2$) and carrier ($T$) lifetimes. The output signal is routed to a computational path for virtual node generation and readout, and to an experimental path for coherence diagnostics via the second-order correlation function $g^{(2)}(0)$.}
  \label{Figure_11}
\end{figure*}

Building on the distinct physical dynamics of the four laser classes, the TDRC architecture fundamentally shifts the traditional neural network paradigm: instead of a spatially distributed network of physical nodes, a single nonlinear node (the laser) with delayed feedback is employed. The implementation is schematically shown in Fig.~\ref{Figure_11}. This time-multiplexing scheme was first introduced by Appeltant et al.~\cite{appeltant2011information} and has been intensively studied since then~\cite{dong2024pssr, wang2024reservoir, hasegawa2023parallel}.

The architecture comprises three primary functional layers: input, reservoir, and output. In the input layer, the temporal input signal \(u(t)\) is sampled and held, then multiplied by a time-multiplexed mask \(M_N\) composed of \(N\) independent random values drawn from a uniform distribution on [-1, 1], each held constant for the duration of the virtual node spacing \(\theta\). In this work, the mask values are drawn from a uniform distribution
over \([-1,1]\), and the input signal is normalized to the same symmetric range, ensuring that no artificial bias is introduced into the pump modulation. This masked signal directly modulates the laser's driving current (pump), thereby injecting the input information into the nonlinear physical reservoir. Importantly, this time-multiplexed masking breaks the symmetry of the input signal, discretizes time, and maps the one-dimensional sequence into a high-dimensional virtual-node state space. 

The masked input is injected into the laser, perturbing its carrier density and driving the optical field into a transient, nonlinear state. The ``reservoir'' is constructed using a delayed optical feedback loop characterized by three control parameters: delay \(\tau\), feedback strength \(K\), and feedback phase \(\phi\). It is worth noting that the present TDRC architecture operates in a non-standard regime in which the feedback delay $\tau$ is not necessarily equal to the mask period $N_v\theta$. In the standard TDRC protocol, one sets $\tau = N_v\theta$ to ensure cyclic coupling between virtual nodes. Here, however, the optimal node spacing $\theta_{\text{opt}}$ is found to scale with the intrinsic relaxation oscillation period $\tau_{\text{ro}}$, which differs by more than an order of magnitude across the four laser classes. Consequently, for a fixed $\tau = 50$ and $N_v = 20$, the matching condition would require $\theta = 2.5$, whereas the optimal values range from $\theta \approx 1.3$ for Class A to $\theta \approx 18.6$ for Class C$_B$. We therefore deliberately allow $\tau \neq N_v\theta$, which generalizes the TDRC concept: the delayed feedback still provides fading memory, but the coupling between virtual nodes becomes local when $\tau < N_v\theta$ and long-range when $\tau > N_v\theta$. This non-standard operation does not invalidate the reservoir computing paradigm; rather, it exploits the intrinsic timescale hierarchy of each laser class. Because the ratio \(\tau/(N_v\theta)\) is no longer fixed, tuning \(\theta\) controls the coupling range between virtual nodes and avoids the degeneracy that arises when the delay is perfectly synchronized with the clock cycle~\cite{hulser2023deriving}: the delayed field represents the laser's past state and couples the current virtual node to previous nodes.

The extraction of processed information occurs via the computational path. The laser's optical output intensity is sampled at intervals \(\theta\), producing a high-dimensional state vector of \(N_v\) virtual nodes per delay period. Only the readout weights \(W_{\text{out}}\) are trained using ridge regression to minimize the prediction error. This makes training exceptionally fast compared to standard recurrent neural networks. 

\subsection{Benchmark Tasks: Mackey-Glass, Lorenz, and Santa Fe}
To quantitatively evaluate the computational capacity, nonlinearity, and memory of the proposed TDRC framework across the four laser classes, we employ three benchmark tasks: the Mackey-Glass (MG) chaotic time series, the three-dimensional Lorenz system, and the Santa Fe laser intensity dataset. For reproducibility, the precise parameters and preprocessing steps of each task are specified below.

\subsubsection{Mackey-Glass Chaotic System}
The MG series exhibits chaotic dynamics in the chosen parameter regime and is a well-known non-Markovian signal, fundamentally driven by a time-delayed nonlinear differential equation~\cite{appeltant2011information}:

\begin{eqnarray}
\frac{dx}{dt} = \beta_{MG} \frac{x(t-\tau_{MG})}{1 + [x(t-\tau_{MG})]^n} - \gamma_{MG} x(t)
\end{eqnarray}

\noindent with the standard chaotic parameters \(\beta_{MG}=0.2\), \(\gamma_{MG}=0.1\), \(\tau_{MG}=17\), and \(n=10\). In this work, we integrate the MG system with a step size of \(\Delta t_{\text{int}} = 0.001\), sample it every \(0.1\) time units, and normalize the input sequence to the range \([-1,1]\). The reservoir is tasked with predicting the state one step ahead, corresponding to a physical time horizon of 0.1 (one sampling interval). This choice ensures that the task demands substantial fading memory capabilities~\cite{appeltant2011information, dong2024pssr}. We emphasize that these parameters differ from those used in earlier literature; therefore, the absolute NRMSE values reported here are not directly comparable to previous benchmark results. Our focus is on the relative performance across the four laser classes under identical conditions.

This task is widely adopted for benchmarking physical reservoirs, as accurate prediction of future values requires the system to maintain fading memory of previous states. This continuous stream is then discretized into time-multiplexed sequences via the input mask \(M_N\), which samples the signal at intervals of \(\theta\) and holds it constant over one virtual-node interval, before being applied to modulate the laser's driving current. The prediction performance is assessed in terms of the Normalized Root Mean Square Error (NRMSE) calculated between the outputs of the linear readout and the ground-truth target sequence.

\subsubsection{Lorenz System}
The Lorenz system is a canonical three-dimensional chaotic flow governed by
\begin{equation}
\dot{x} = \sigma(y - x),\quad \dot{y} = x(\rho - z) - y,\quad \dot{z} = xy - \beta z,
\label{eq:lorenz}
\end{equation}
with the standard parameters $\sigma = 10$, $\rho = 28$, and $\beta = 8/3$. The system is integrated with a fourth-order Runge--Kutta scheme using a step size of $\Delta t_\mathrm{int} = 0.001$, and the trajectory is sampled every $0.02$ time units. The $x$-component is linearly normalized to $[-1, 1]$. The reservoir is tasked with predicting the state two steps ahead, corresponding to a physical horizon of 0.04. Compared with the MG task, the Lorenz system involves three coupled degrees of freedom and a more complex attractor topology, imposing a stronger demand on the reservoir's nonlinear mapping capacity.

\subsubsection{Santa Fe Laser Intensity Dataset}
We use Dataset A from the 1994 Santa Fe Time Series Prediction Competition~\cite{Hubner1989, weigend1994time}, which records the intensity of a far-infrared laser operating in a chaotic regime. The distributed record contains 10093 consecutive samples; the sequence is linearly normalized to \([-1, 1]\) without further preprocessing. The reservoir is tasked with predicting the intensity two sampling steps ahead, where one step corresponds to one consecutive data point. Because the experimental record does not specify a physical sampling interval, the prediction horizon is expressed in units of data points rather than in physical time. Because the dataset originates from an experimental single-mode laser system, it provides a realistic test of the reservoir's ability to handle noise and non-ideal dynamics. 

\subsection{Numerical Simulation and System Parameters}
The TDRC system is modelled using the complete Maxwell-Bloch equations with a 4th-order Runge-Kutta (RK4) integration scheme. Spontaneous emission noise is added as an additive white Gaussian term after each RK4 step~\cite{Pyragas2025}. To isolate performance differences caused by internal laser physics, the external macroscopic delay \(\tau\) was kept identical across all classes (50 units of \(T_{\text{ph}}\)). A fixed integration step \(dt = 0.002\) was uniformly adopted across all laser classes, which remains sufficiently small to guarantee numerical stability and precise temporal alignment even for the fastest internal dynamics (e.g., \(T_2 = 0.01\)). The intrinsic physical parameters governing the laser dynamics and the numerical integration settings are summarized in Table~\ref{tab:phys_params}.

\begin{table*}[htbp]
\centering
\renewcommand{\arraystretch}{1.5}
\caption{Intrinsic physical parameters and numerical simulation settings for the four laser classes. All time values are normalized to the cavity photon lifetime \(T_{\text{ph}}\).}
\label{tab:phys_params}
\begin{tabular}{|l|c|c|c|c|c|}
\hline
\textbf{Parameter} & \textbf{Symbol} & \textbf{Class A} & \textbf{Class B} & \textbf{Class C\(_A\)} & \textbf{Class C\(_B\)} \\
\hline
Polarization lifetime & \(T_2\) & 0.01 & 0.01 & 2.0 & 2.0 \\
\hline
Carrier lifetime & \(T\) & 0.01 & 392.0 & 2.0 & 392.0 \\
\hline
Normalized cavity decay rate & \(c\) & \multicolumn{4}{c|}{\((\frac{2\delta\omega}{T_2+2})^2+1\)} \\
\hline
Physical feedback delay & \(\tau\) & \multicolumn{4}{c|}{50.0} \\
\hline
Number of virtual nodes & \(N_v\) & \multicolumn{4}{c|}{20} \\
\hline
Integration time step & \(dt\) & \multicolumn{4}{c|}{0.002} \\
\hline
Integration scheme & & \multicolumn{4}{c|}{4th-order Runge-Kutta} \\
\hline
\end{tabular}
\vspace{0.2cm}
\parbox{\linewidth}{\footnotesize The parameter \(c\) ensures the first laser threshold at \(P_{\text{pump}}=0.5\). The optimized control parameters (\(\kappa\), \(P_{base}\), \(\theta\), \(\gamma\), \(\delta\omega\), \(\phi\)) are summarized separately in Table 2. The intermediate quantum-dot-like Class A\(^Q\) (\(T_2=0.01\), \(T=2.0\)) is documented in the Supplementary Information.}
\end{table*}

\subsection{Fair Comparison Methodology}
To guarantee a fair comparison among the four laser classes, we adhere to three methodological principles. First, steady-state intensity normalization: the pump bias \(P_{\mathrm{base}}\) for each class is uniformly set to \(P_{\mathrm{base}} = 1.95\), establishing a consistent baseline signal-to-noise ratio. Second, individual parameter optimization: the node spacing \(\theta\) and input scaling factor \(\gamma\) are separately scanned and optimized for each class via high-resolution NRMSE grid searches, so that the subsequent comparison captures the best achievable performance of each laser class rather than the performance evaluated at an arbitrarily chosen shared parameter set. Third, unified data protocol: for all three benchmarks, the first \(3{,}000\) samples are used for training and the subsequent \(5{,}000\) for testing, with a washout of 100 samples at the beginning of each sequence.

To avoid temporal degeneracy and to reflect realistic reservoir operation, the number of virtual nodes is set to \(N_v = 20\) for all classes. Then, to understand the dynamical origin of the computational differences among the four laser classes, we investigate the intrinsic temporal responses of each laser type to a transient pump perturbation. The distinct response characteristics, from rapid monotonic recovery in Class A, damped relaxation oscillations in Class B, to high-frequency coherent ringing enabled by the finite polarization lifetime in Class \(C_A\) and Class \(C_B\) lasers, are shown in the Supplementary Material (Supplementary Fig.~S1). These underlying dynamical responses fundamentally shape the fading memory and nonlinearity available for the TDRC tasks.

\section{Results and Discussion}
\subsection{Physical Phase Diagrams and Topological Time-Scale Signatures}
\begin{figure*}[ht!]
\centering
  \includegraphics[width=15.0cm, height=6.5cm]{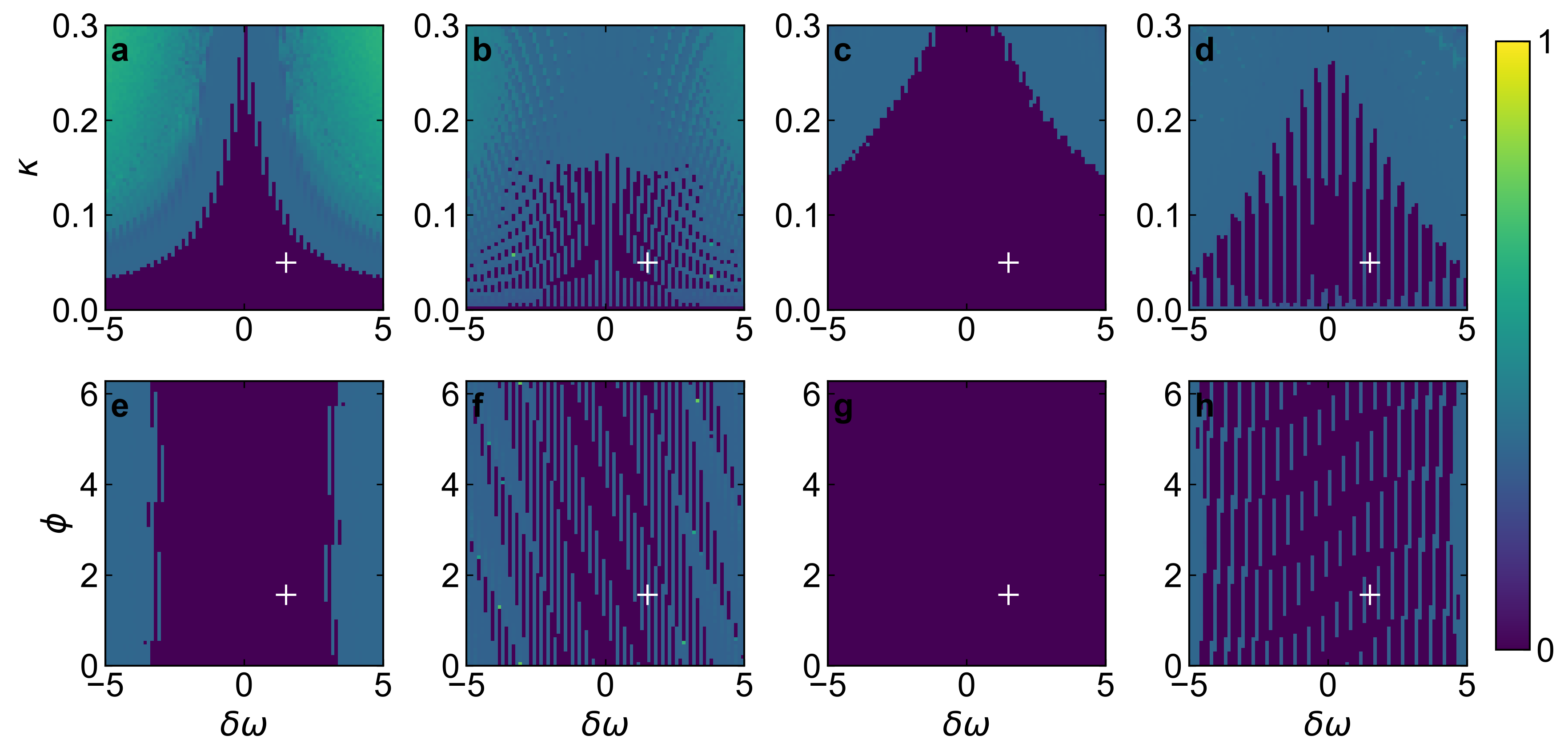}
  \caption{Permutation Entropy (PE) mappings of the four laser classes in two distinct parameter spaces: (a--d) the feedback strength--detuning plane $(\delta\omega, \kappa)$ at fixed phase \(\phi=\pi/2\) and (e--h) the feedback phase--detuning plane $(\delta\omega, \phi)$ at fixed feedback strength
  \(\kappa=0.05\). Panels (a, e) represent Class A; (b, f) Class B; (c, g) Class C\(_A\); and (d, h) Class C\(_B\). The white cross in every panel marks the unified physical baseline \((\delta\omega=1.5,\ \kappa=0.05,\ \phi=\pi/2)\) used for all task evaluations in this work; its position is \((1.5,0.05)\) in the \((\delta\omega,\kappa)\) plane and \((1.5,\pi/2)\) in the \((\delta\omega,\phi)\) plane.}
  \label{PE_Bifurcation}
\end{figure*}

To investigate how the interplay between polarization and carrier dynamics governs the operational phase space, we carry out extensive parameter sweeps and use Permutation Entropy (PE) to quantify dynamical complexity~\cite{Gancio2025}. As shown in Fig.~\ref{PE_Bifurcation}, the PE mappings for the four laser classes are mapped across two distinct parameter spaces: the feedback strength--detuning landscape ($\delta \omega$, $\kappa$) (Figs.~\ref{PE_Bifurcation}a--d) and the feedback phase--detuning landscape ($\delta \omega$, $\phi$) (Figs.~\ref{PE_Bifurcation}e--h). In these color maps, deep purple regions indicate low entropy (stable continuous-wave operation or highly regular periodic limit cycles), while the transition toward green and yellow represents elevated entropy, indicative of complex chaotic dynamics.

We first examine the $(\delta\omega, \kappa)$ parameter space. Class A and Class C$_A$ exhibit smooth, parabolic continuous-wave (CW) stability basins, characteristic of strongly damped systems where fast relaxation suppresses long-term memory accumulation~\cite{virte2013deterministic, takemura2021low}. Class B (Fig.~\ref{PE_Bifurcation}b), by contrast, reveals a highly structured network of multistable locking domains, a well-known signature of the intricate bifurcation structure in delayed feedback lasers~\cite{wieczorek2005dynamical, Marconi2020, sciamanna2015physics}. This deterministic structural confinement, arising from the beating between relaxation oscillations and the external cavity frequencies, which are spaced by ($1/\tau$)~\cite{Marconi2020}, restricts the edge-of-stability dynamics to narrow parameter islands~\cite{appeltant2011information}. In contrast, Class C$_B$ (Fig.~\ref{PE_Bifurcation}d) demonstrates the topological smoothing power of a finite polarization lifetime. Acting as an intrinsic low-pass filter~\cite{haken1984laser, van1998laser}, it effectively dissolves the sharp locking boundaries of Class B into a broadened, continuous, and parametrically resilient operational basin.

We next evaluate the $(\delta\omega, \phi)$ parameter space. As shown in Figs.~\ref{PE_Bifurcation}e and \ref{PE_Bifurcation}g, both Class A and Class C$_A$ display vertically uniform, low-entropy backgrounds, which in both cases stems from the suppression of phase interference due to either fast damping (Class A) or short-term memory (Class C$_A$). 

Class B (Fig.~\ref{PE_Bifurcation}f), by contrast, displays a striking pattern of diagonally striated, discrete ``short-bar-like'' structures, which are governed by the phase-matching condition $\delta\omega\cdot \tau + \phi = 2m\pi$~\cite{wunsche2005synchronization}. In contrast, Class C$_B$ (Fig.~\ref{PE_Bifurcation}h) smears these discrete bars into continuous, broadened diagonal bands, once again demonstrating the smoothing effect of its finite polarization lifetime as a low-pass filter~\cite{romeira2017delay}. 

These distinct topologies, as illustrated in Supplementary Fig. S1, are fundamentally rooted in the lasers' internal transient dynamics and validate our central thesis: the slow polarization of Class C$_B$ acts as a topological smoother, dissolving the sharp bifurcation boundaries and phase sensitivity observed in the strongly damped and long-memory regimes, thus unifying hardware-robust stability with the rich dynamics required for photonic neuromorphic processors.

Based on these diagrams, we choose a unified detuning of $\delta \omega = 1.5$ for all subsequent task evaluations. This choice is motivated by three considerations. First, it lies in a region where all four laser classes exhibit non-trivial dynamics-neither deep in the stable continuous-wave regime nor beyond the edge of stability-thereby allowing each class to operate near its optimal computational regime. Second, it is sufficiently far from the dense multi-stable locking domains of Class B to avoid the fragile ECM locking that would otherwise dominate, while still preserving the topological smoothing effect in Class C$_B$. Third, it provides a common ground for fair comparison, ensuring that the observed performance differences arise from the intrinsic time-scale hierarchies rather than from arbitrary detuning choices. This detuning is therefore adopted as the unified physical baseline in Table II and used throughout the task evaluations. This baseline is marked by a white cross in every panel of Fig.~\ref{PE_Bifurcation}. The subsequent optimization of task-specific parameters ($\theta$, $\gamma$) is then conducted on this common physical foundation.

\subsection{Optimal Parameter Mapping Guided by Task-Driven Performance}
To establish an unbiased physical platform for comparing the four laser classes, we first map the global predictive performance across the pump-feedback parameter space ($P_{base}, \kappa$), with a detailed phenomenological analysis provided in the Supplementary Materials (Fig. S2). Briefly, the error landscapes reveal that the ultra-long carrier memory of Class B restricts stable operation to narrow parameter corridors bounded by horizontal resonant bands, while Class C\(_A\) suffers from localized optical quenching (i.e., gain truncation) under simultaneous high-pump and strong-feedback driving. In contrast, Class C\(_B\) utilizes its finite polarization lifetime ($T_2 = 2.0$) as a dynamic low-pass filter to maintain an expansive, remarkably uniform stability plateau. Guided by these macroscopic topological boundaries, we establish a universally stable baseline at $P_{base} = 1.95$ and $\kappa = 0.05$ (weak-feedback regime). This unified operating point safely navigates the dynamical constraints of Class B and the over-saturation limits of Class C\(_A\), ensuring that the subsequent optimization of task-specific information encoding parameters (virtual node interval $\theta$ and input scaling factor $\gamma$) is conducted on a rigorously equitable physical foundation.

It is necessary to mention that this baseline defines only the intrinsic laser physics; the task-specific information-encoding parameters ($\theta$ and $\gamma$) are independently optimized for each class in a two-dimensional grid search to capture their respective maximum computational capacity. This systematic optimization strategy follows established best practices for photonic RC parameter tuning and TDRC operating-point selection~\cite{brunner2013parallel, jaeger2004harnessing, dong2024pssr}.

We perform high-resolution two-dimensional ($\theta$, $\gamma$) scans for each laser class under the unified physical baseline. The resulting optimal operating points are summarized in Table II. The 2D grid search locates physically implementable optima for all classes, with $\gamma$ remaining within a normal driving range ($0.28 \le \gamma \le 1.39$). This range corresponds to the standard weak-to-moderate modulation regime in which the laser operates below the gain-truncation boundary, ensuring that the optimized operation points are compatible with physically realizable experimental conditions.

\begin{figure*}[ht!]
\centering
\includegraphics[width=15.0cm, height=7.2cm]{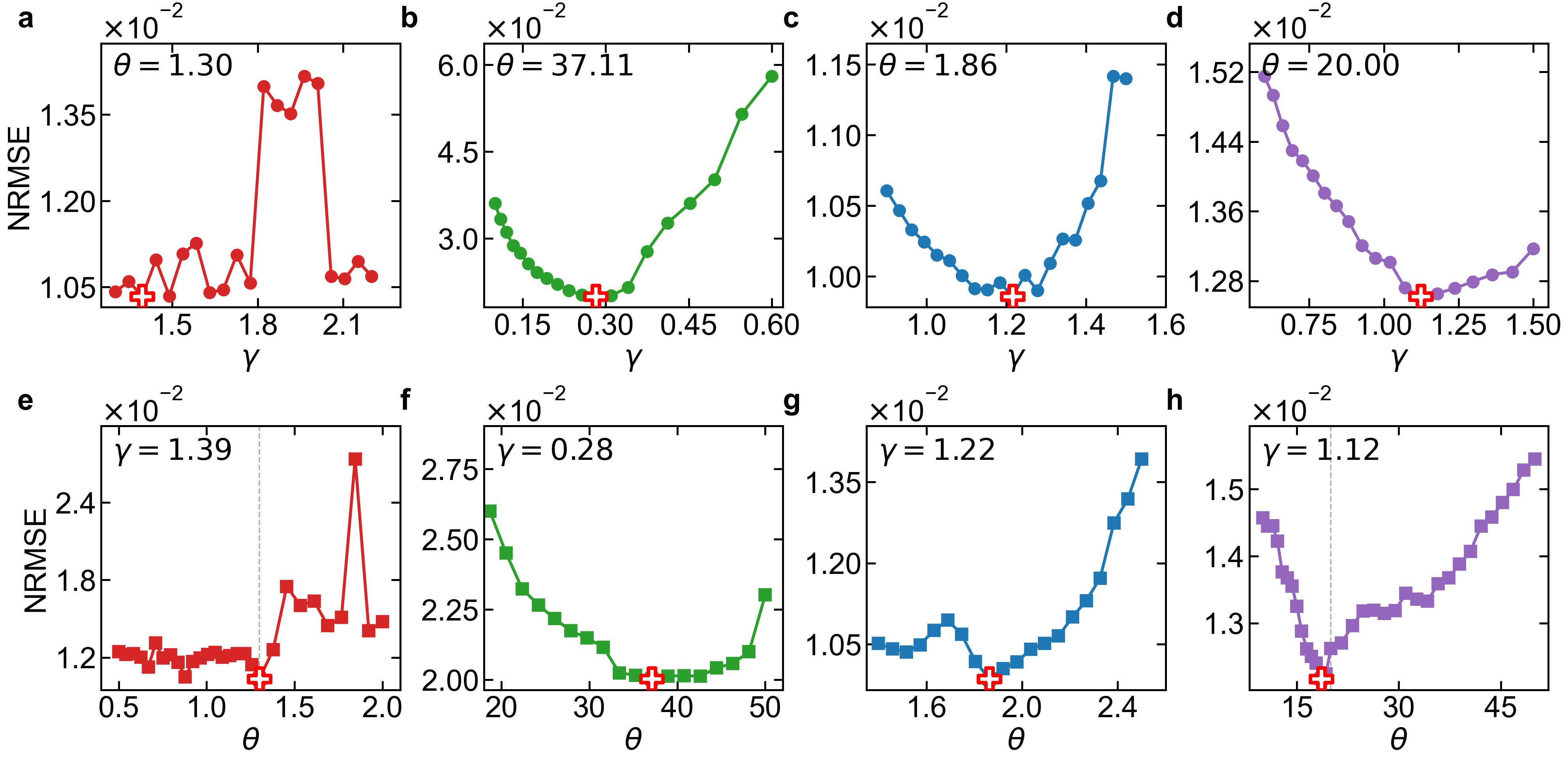}
\caption{High-resolution 1D parameter sweeps for the four primary laser classes. Top row (a-d): NRMSE as a function of the input scaling factor 
$\gamma$ at each class's optimal node interval $\theta_{\mathrm{opt}}$. Bottom row (e–h): NRMSE as a function of the virtual node interval $\theta$ at each class's optimal $\gamma_{\mathrm{opt}}$. Columns correspond to Class A (a, e), Class B (b, f), Class C$_A$ (c, g), and Class C$_B$ (d, h). Each panel shows a single class, color-coded as follows: Class A in red, Class B in green, Class C$_A$ in blue, and Class C$_B$ in purple. The red P-shaped markers denote the globally optimal parameters yielding the minimum NRMSE. In panels (e) and (h), the dashed vertical line marks the boundary between the main scan and the high-resolution supplementary scan for Class A and Class C$_B$.}
\label{NMSE_Scan_Theta_Gamma}
\end{figure*} 

The resulting optimization landscapes exhibit distinct physical signatures directly reflecting each laser's temporal-scale hierarchy. For Class A (Fig.~\ref{NMSE_Scan_Theta_Gamma}a, e), the $\gamma$-scan at their respective optimal $\theta$ reveals smooth, broad valleys with the minimum located at $\gamma \approx 1.4$. This operating point works for the chosen parameters for all the 4 lasers and they do not transition into chaotic behaviour. The $\theta$-scan exhibits a broad but shallow valley, confirming the absence of an intrinsic resonance and consistent with the memoryless nature of these classes. 

For the Class B lasers (Fig.~\ref{NMSE_Scan_Theta_Gamma}b, f), the $\theta$-scan reveals a pronounced V-shaped valley, with an optimum located at $\theta_{\text{opt}} = 37.11$. This significantly longer node interval, corresponding to $\theta_{opt}/\tau_{ro} \approx 1.32$, allows the ultra-long carrier lifetime ($T = 392.0$) to accumulate temporal correlations without degenerating into a static-mapping regime. The corresponding $\gamma$-scan is dominated by an extremely shallow optimum at $\gamma_{opt} = 0.28$, indicating that Class B requires a very weak input coupling to avoid over-driving its slow carrier reservoir.

For Class C$_A$ (Fig.~\ref{NMSE_Scan_Theta_Gamma}c, g), with $T_2 = T = 2.0$, the optimal operating point is at $\theta_{opt}=1.86$ and $\gamma_{opt} = 1.22$. The corresponding ratio $\theta_{opt}/\tau_{ro} \approx 0.93$ places this laser squarely within the classical dynamic TDRC regime. Both the $\gamma$-scan and the $\theta$-scan exhibit smooth, well-defined parabolic minima, confirming that Class C$_A$ operates at a genuinely balanced operating point where both carrier and polarization dynamics contribute constructively.

In contrast to Class B, the Class C$_B$ laser (Fig.~\ref{NMSE_Scan_Theta_Gamma}d, h) forms a distinctly different landscape. Its $\theta$-scan (obtained by combining the main scan $\theta \in [20, 50]$ with a high-resolution supplementary scan $\theta \in [10, 20]$) exhibits a minimum value around $\theta_{opt}=18.57$ ($\theta_{opt}/\tau_{ro} \approx 0.66$). This pronounced sensitivity to $\theta$ detuning is a direct manifestation of the slow polarization lifetime ($T_2 = 2.0$). By acting as an inline temporal filter, polarization dynamics smooth out high-frequency inter-node timing mismatches, which shifts the optimal node spacing relative to Class B. The $\gamma$-scan locates the optimum at $\gamma_{opt}=1.12$, which is comparable in magnitude to that of Class C$_A$, demonstrating that the polarization-induced low-pass filtering operates at a fundamentally similar driving regime as the perfectly matched Class C$_A$.

These optimization results demonstrate that engineering parameters cannot be chosen arbitrarily. They must be precisely aligned with the intrinsic relaxation frequencies and dynamic damping rates of the underlying physical substrate. In particular, the physically implementable operating points identified here (all with $\gamma \leq 1.4$) remain well below the gain-truncation boundary that limits the memoryless classes. The perfect time-scale matching of Class C$_A$ provides the most energy-efficient operation, and the low-pass filtering of Class C$_B$ offers the broadest operational tolerance. Table~\ref{tab:optimized_params} summarizes the unified physical baseline alongside the optimal control parameters across all four laser classes.

\begin{table*}[htbp]
\centering
\renewcommand{\arraystretch}{1.4}
\caption{Summary of the unified physical baseline and the optimized control parameters derived from the NRMSE scans.}
\label{tab:optimized_params}
\begin{tabular}{|l|c|c|c|c|c|c|}
\hline
\textbf{Laser Class} & \multicolumn{4}{c|}{\textbf{Unified Physical Baseline}} & \multicolumn{2}{c|}{\textbf{Optimized Engineering Params.}} \\
\cline{2-7}
 & \(\kappa\) & \(P_{base}\) & \(\delta\omega\) & \(\phi\) & \(\theta_{\text{opt}}\) & \(\gamma_{\text{opt}}\) \\
\hline
Class A & 0.05 & 1.95 & 1.5 & \(\pi/2\) & 1.30 & 1.39 \\
\hline
Class B & 0.05 & 1.95 & 1.5 & \(\pi/2\) & 37.11 & 0.28 \\
\hline
Class C\(_A\) & 0.05 & 1.95 & 1.5 & \(\pi/2\) & 1.86 & 1.22 \\
\hline
Class C\(_B\) & 0.05 & 1.95 & 1.5 & \(\pi/2\) & 18.57 & 1.12 \\
\hline
\end{tabular}
\vspace{0.2cm}
\end{table*}

\subsection{High-Precision Prediction and Multi-Mask Robustness}
Having established the optimal physical baseline and engineering parameters (Table~\ref{tab:optimized_params}), we evaluate the high-precision predictive performance and stochastic encoding robustness across the four primary laser classes. The reservoir models were trained on 3,000 consecutive data points and evaluated on an independent out-of-sample test set comprising 5,000 time steps of the Mackey-Glass chaotic time series. To quantitatively benchmark input encoding resilience, we performed 20 independent evaluation trials for each laser class, employing a newly generated, distinct random continuous uniform in each iteration.

\begin{figure}[ht!]
\centering
\includegraphics[width=8.0cm, height=3.8cm]{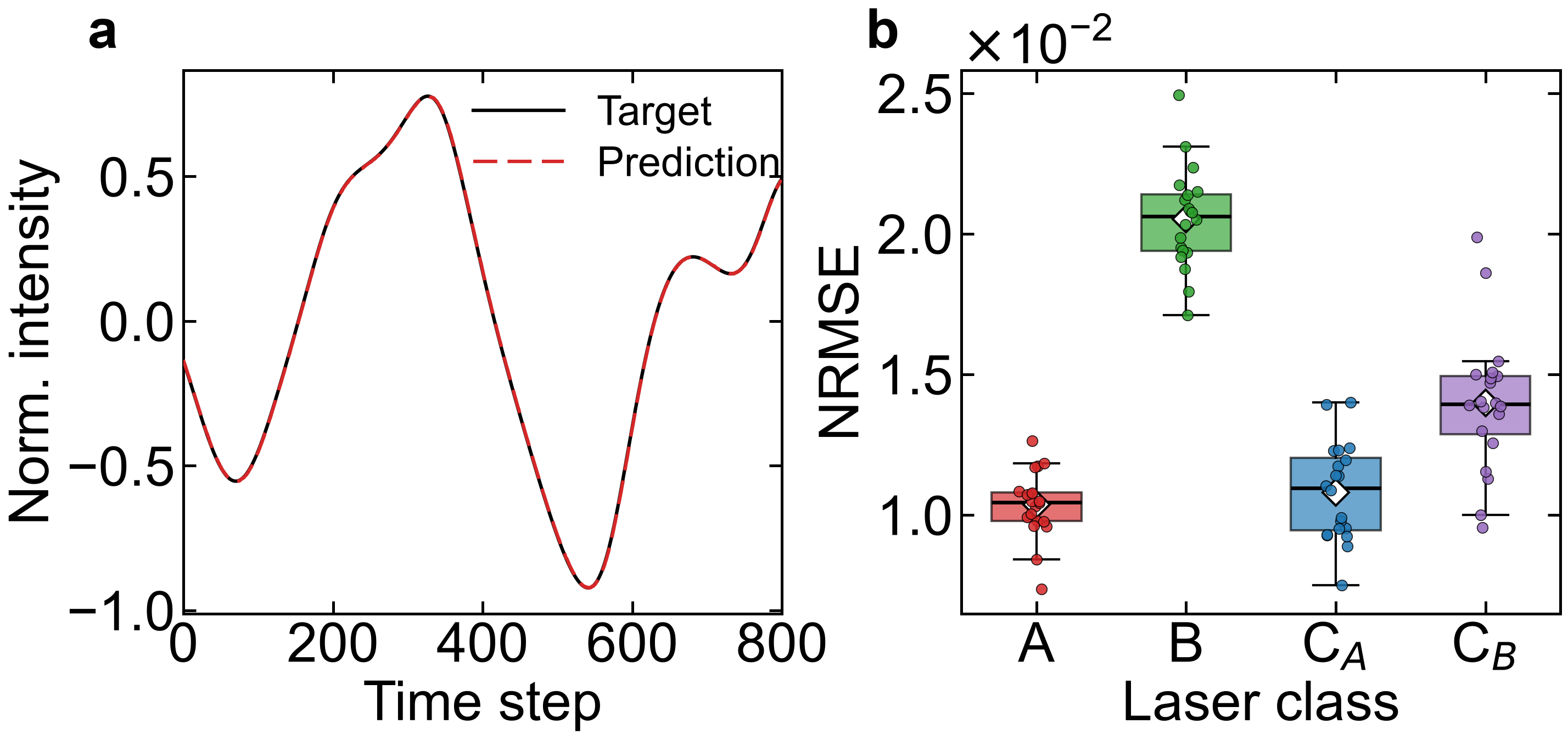}
\caption{Multi-mask robustness. (a) Representative 800-step prediction trajectory. The thin black curve is the true target, the thick red dashed curve is the ensemble average over 20 independent random masks, and the shaded region indicates the $\pm$1$\sigma$ band. The prediction closely follows the target, and the narrow $\pm$1$\sigma$ band indicates small mask-to-mask variance. (b) Statistical distributions of the NRMSE over the 20 mask realizations for all four laser classes. Boxes show the interquartile range, black lines the median, white diamonds the mean, and colored dots the individual mask realizations. The similar box widths indicate that the mask-induced variance is comparable across classes, while the different box positions reflect the distinct intrinsic nonlinearity of each laser class. The performance differences reported in Table III therefore originate from the intrinsic laser physics rather than from mask randomness.}
\label{Task_Robust_20_mask}
\end{figure}

Fig.~\ref{Task_Robust_20_mask} summarizes the multi-mask prediction performance. Fig.~\ref{Task_Robust_20_mask}a shows a representative 800-step prediction trajectory for Class A. The thin black curve is the true target, the red dashed curve is the ensemble average over 20 independent random masks, and the shaded region denotes the $\pm$1$\sigma$. The prediction closely follows the target, and the narrow $\pm$1$\sigma$ band indicates small mask-to-mask variance, confirming that the optimized parameter configuration provides the complex transient response required for this task.

Fig.~\ref{Task_Robust_20_mask}b compares the NRMSE distributions across the 20 mask realizations for all four laser classes. The distributions are tight and of comparable width for all classes (standard deviations between 0.001 and 0.002), indicating that the mask-to-mask variation is consistent across classes and does not alter their relative ranking. The mean NRMSE values, however, differ systematically: Class A achieves 0.010, Class C$_A$ 0.011, Class C$_B$ 0.014, and Class B 0.021. These differences reflect the distinct intrinsic nonlinearity and time-scale hierarchy of each class: the instantaneous response of Class A supports a stable static nonlinear mapping; the perfectly matched time scales of Class C$_A$ yield the richest intermediate-order nonlinear response; the finite polarization lifetime of Class C$_B$ acts as an intrinsic low-pass filter that slightly smooths high-frequency mask components; and the ultra-long carrier lifetime of Class B limits its response speed. The performance differences reported in Table III therefore originate from the intrinsic laser physics rather than from mask randomness.
  
The ability to sustain low prediction error with minimal variance across stochastic mask realizations provides an essential physical foundation for practical, hardware-ready neuromorphic processors~\cite{goldmann2020deep}.

\subsection{Task Generalization and Information Processing Capacity Decomposition}
To assess whether the observed performance trends are specific to a single benchmark or reflect a more general task dependence, we evaluate the four primary laser classes against two additional, distinct chaotic benchmarks: the three-dimensional Lorenz system and the experimental Santa Fe laser intensity dataset. All evaluations are performed under the unified physical baseline and class-optimized control parameters established in Table~\ref{tab:optimized_params}. For each task, the readout weights are retrained on the corresponding training set. Table~\ref{tab:generalization} summarizes the cross-task prediction performance, reported as the mean Normalized Root Mean Square Error ($\text{NRMSE}$) alongside its standard deviation ($\pm 1\sigma$) across 10 independent stochastic mask realizations.

\begin{table*}[htbp]
\centering
\renewcommand{\arraystretch}{1.5}
\caption{Task Generalization Performance (NRMSE \(\pm\) Standard Deviation)
across different benchmark tasks. All values are the mean \(\pm\) standard
deviation over 10 independent random masks.}
\label{tab:generalization}
\begin{tabular}{|l|c|c|c|}
\hline
\textbf{Laser Class} & \textbf{Mackey-Glass} & \textbf{Lorenz} &
\textbf{Santa Fe (Dataset A)} \\
\hline
Class A & \(0.0103 \pm 0.0014\) & \(0.0567 \pm 0.0148\) &
\(0.5233 \pm 0.1631\) \\
\hline
Class B & \(0.0217 \pm 0.0031\) & \(0.1544 \pm 0.0400\) &
\(0.6955 \pm 0.0426\) \\
\hline
Class C\(_A\) & \(0.0105 \pm 0.0024\) & \(0.0506 \pm 0.0142\) &
\(0.4251 \pm 0.0651\) \\
\hline
Class C\(_B\) & \(0.0141 \pm 0.0013\) & \(0.0763 \pm 0.0321\) &
\(0.5341 \pm 0.0650\) \\
\hline
\end{tabular}
\end{table*}

The empirical benchmarks reveal a strongly task-dependent performance landscape, shaped by the interplay between driving strength and intrinsic time scales. On the short-horizon Mackey--Glass task, Class A (0.0103) and Class C\(_A\) (0.0105) achieve comparable, and the lowest, NRMSE---their difference is well within one standard deviation---while Class A exhibits a smaller mask-to-mask spread (0.0014 vs.\ 0.0024). This result is consistent with the operation of both classes in the static-nonlinear-mapping regime: their high content of instantaneous response is ideally suited to this essentially short-memory task. However, as the task demands higher-order nonlinear complexity over a longer effective horizon, the matched time scales of Class C\(_A\) (\(\theta_\mathrm{opt}/\tau_\mathrm{ro} \approx 0.93\)) become decisive, and Class C\(_A\) overtakes Class A on both the Lorenz (0.0506 vs.\ 0.0567) and the Santa Fe (0.4251 vs.\ 0.5233) benchmarks. The advantage of Class C\(_A\) is particularly pronounced on the experimental Santa Fe
dataset, where it leads Class A by roughly 20\%, whereas the same comparison on the deterministic Lorenz system yields a narrower margin of
about 11\%. This distinction suggests that real-world data, with its non-stationarity and measurement noise, rewards the balanced time-scale
hierarchy of Class C\(_A\) even more strongly than idealized chaotic flows. Notably, the mask-to-mask spread of Class A on Santa Fe
($\pm 0.1631$) is more than twice that of Class C\(_A\) ($\pm 0.0651$), indicating that the matched time scales of Class C\(_A\) also provide a substantially more robust operating point for noisy experimental data.

Class B exhibits the poorest performance across all three benchmarks (MG: \(0.0217\), Lorenz: \(0.1544\), Santa Fe: \(0.6955\)). The ultra-long carrier lifetime (\(T = 392.0\)) combined with the polarization-free dynamics makes the system respond too slowly for the rapidly varying Lorenz and Santa Fe attractors, and the weak input coupling required to stabilize the slow carrier reservoir limits the amount of information that can be encoded.

In contrast, Class C\(_B\) occupies an intermediate position, outperforming Class B by 35\% on Mackey--Glass (0.0141 vs 0.0217), 51\% on Lorenz (0.0763 vs 0.1544), and 23\% on Santa Fe (0.5341 vs 0.6955). The polarization-mediated low-pass filtering plays a decisive role: by smoothing the high-frequency components of the deterministic input mask and enabling a stronger input coupling (\(\gamma_\mathrm{opt} = 1.12\) vs.\ \(0.28\) for Class B), Class C\(_B\) simultaneously achieves a larger effective modulation depth and a richer nonlinear feature space. The task-dependent improvement is largest for the Lorenz benchmark, which combines a longer prediction horizon with a deterministic, noise-free attractor; on the experimental Santa Fe dataset, the additional noise partially offsets the benefit of the polarization filter, and the improvement over Class B is correspondingly smaller. On Santa Fe, Class C\(_B\) (0.5341) and Class A (0.5233) are statistically indistinguishable, as their difference is much smaller than the standard deviation of either class; nevertheless, Class C\(_B\) retains a systematic advantage over Class B on all three benchmarks, confirming the role of the finite polarization lifetime as a nonlinear-capacity booster.

These results highlight a fundamental trade-off. Short-horizon nonlinear mapping favors fast carriers (Class A), while long-horizon chaotic prediction requires the balanced time scales of Class C\(_A\). The optimal architecture is therefore not a fixed hardware choice but a task-adaptive design decision.

It is important to note that the absolute NRMSE values reported in Table~\ref{tab:generalization} are not directly comparable to those in
the literature. Most photonic TDRC demonstrations of Mackey--Glass prediction employ substantially more virtual nodes (typically \(N_{\nu} \geq 50\), often several hundred) and different prediction horizons~\cite{appeltant2011information, brunner2013parallel, Larger2017, dong2024pssr}. In contrast, the present work uses only \(N_{\nu} = 20\) and a relatively short horizon of 0.1 for Mackey--Glass. Under these more constrained conditions, the NRMSE values of 0.010--0.022 for Mackey--Glass and 0.051--0.154 for Lorenz fall within the range reported for photonic reservoir computing
platforms. The absolute NRMSE on the Santa Fe dataset (0.425--0.696) is substantially larger, as expected for a real experimental record with non-stationary dynamics and measurement noise; values in the range 0.3--0.6 are typical for this benchmark in the literature. Across all three tasks, the central result of this work is not the absolute accuracy but the relative ranking of the four laser classes, which remains consistent with the IPC decomposition presented below.

To systematically unravel the physical origin of these variations, we employ the Information Processing Capacity (IPC) framework based on orthogonal Legendre polynomial expansions. The IPC metric, originally developed by Dambre et al. \cite{dambre2012scientific} and further refined with finite-sample bias correction by Kubota et al., provides a task-independent measure of a system's linear memory and nonlinear computational capability \cite{kubota2021physical}. The degree-decomposed IPC spectrum provides a quantitative mapping of linear memory versus higher-order nonlinear transformation capacity. Recent studies have successfully demonstrated the effectiveness of this decomposition in designing optimized photonic RC systems \cite{huang2025stochastic, hulser2023deriving}.

\begin{figure}[ht!]
\centering
\includegraphics[width=7.0cm, height=6.0cm]{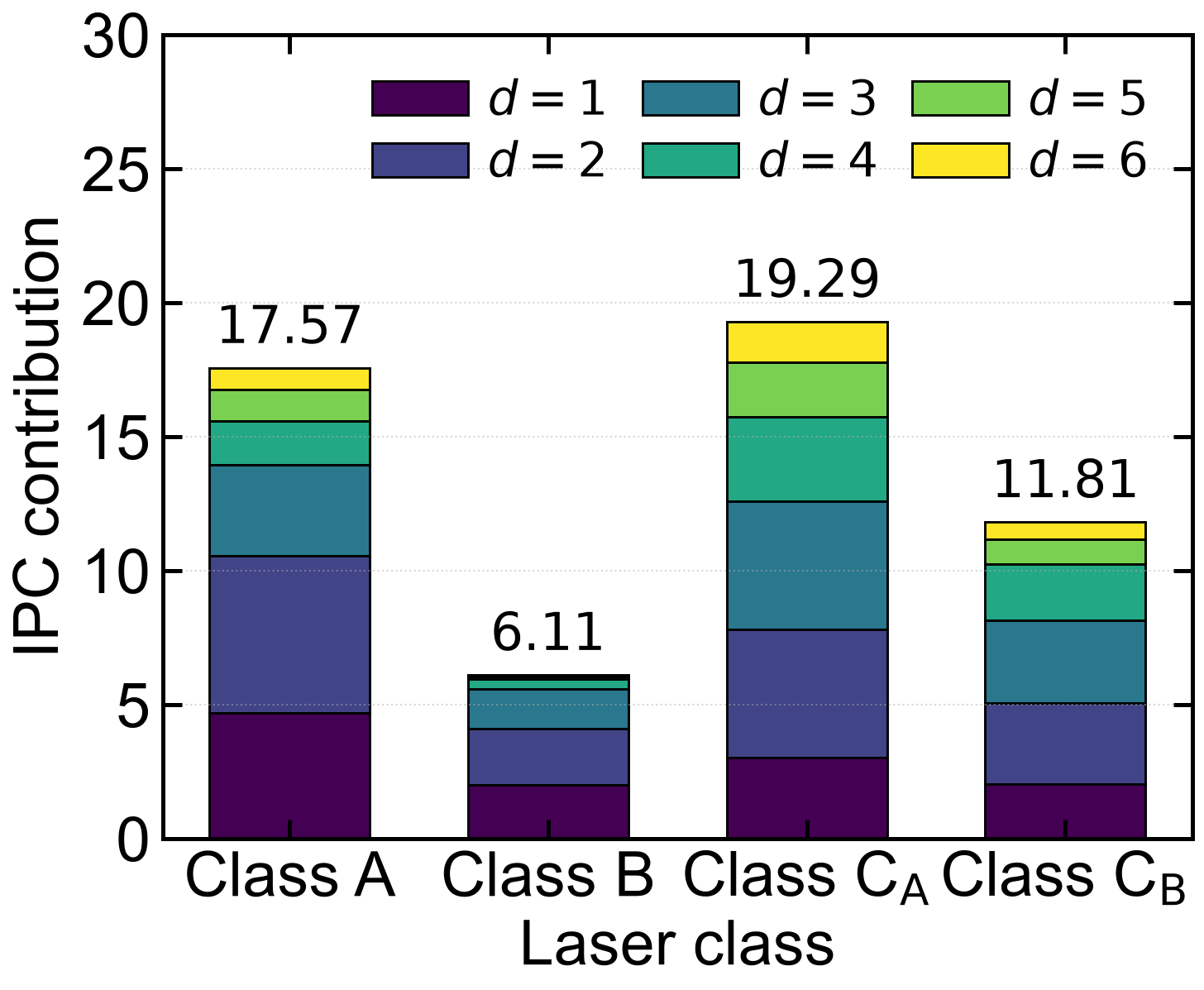}
\caption{Information Processing Capacity (IPC) decomposition across the four primary laser classes, evaluated at each class's optimal control 
parameters (Table II). The stacked bars show the spectral distribution of computational capacity across Legendre polynomial degrees from 
$d = 1$ (linear memory) to $d = 6$ (high-order nonlinearity). The numbers above each bar ($\Sigma C$) indicate the total capacity, i.e., the sum over all degrees $d = 1$–$6$. The x-axis labels the four laser classes; the y-axis is the IPC contribution.}
\label{IPC_stacked_bar_Lasers}
\end{figure}

The IPC decomposition (Fig.~\ref{IPC_stacked_bar_Lasers}) reveals a nuanced hierarchy that, when interpreted together with the benchmark results, quantitatively explains the task-dependent reversal observed in Table~\ref{tab:generalization}. The four laser classes display markedly different linear vs.\ nonlinear capacity balances.

The Class A laser achieves the highest linear IPC ($4.70$), reflecting its instantaneous response. Its nonlinear IPC ($12.87$) is substantial but lower than that of Class C$_A$. This specific balance -- high linear, moderate nonlinear -- perfectly matches the requirements of short-horizon Mackey-Glass prediction.

The Class B laser exhibits the lowest total capacity across both linear ($1.99$) and nonlinear ($4.11$) components. The ultra-long carrier lifetime severely restricts the achievable number of independent polynomial mappings, consistent with the systematically poor performance of this class on all three benchmarks.

Class C$_A$ achieves the highest nonlinear IPC ($16.28$) and the second-highest total capacity ($19.29$). The dominant nonlinear contribution reflects the capacity of the well-matched time scales to generate a rich hierarchy of higher-order Legendre features --- precisely the resource that the Lorenz and Santa Fe tasks require.

Class C$_B$ displays a substantial nonlinear IPC ($9.78$) that is $2.4\times$ larger than that of Class B, while its linear IPC ($2.03$) remains comparable. This selective enhancement of nonlinear capacity directly underpins the 52\% improvement over Class B on the Lorenz and Santa Fe tasks. Notably, the polarization filter acts purely as a nonlinear-capacity booster; it does not affect the linear component, which is why Class C$_B$'s advantage over Class B is much larger for the higher-dimensional chaotic benchmarks than for the short-horizon Mackey-Glass task.

The correspondence between IPC and benchmark performance is striking. The ranking of the four classes by linear IPC (A $>$ C$_A$ $>$ C$_B$ $>$ B) closely matches the ranking by Mackey-Glass accuracy, with Class A and Class C$_A$ statistically indistinguishable. The ranking by nonlinear IPC (C$_A$ $>$ A $>$ C$_B$ $>$ B) matches the Lorenz ranking exactly and the Santa Fe ranking to within the mask-to-mask uncertainty, where Class A and Class C$_B$ are statistically indistinguishable. This agreement demonstrates that the IPC framework provides a rigorous physical basis for understanding the task-dependent performance of the laser-based TDRC systems.

These generalization benchmarks, directly illuminated by IPC spectral decomposition, confirm that carrier and polarization relaxation lifetimes do not merely dictate total capacity magnitudes; they fundamentally sculpt the system's intrinsic computational spectrum. The optimal architecture must achieve a balance between memory depth and responsiveness that matches the specific dynamical demands of the target task. In the present benchmark suite, this balance is best realized by Class C$_A$ for the nonlinear chaotic tasks and by Class A for the short-horizon prediction task, whereas the extreme underdamping of Class B and the (partial) over-smoothing of Class C$_B$ lead to systematically reduced performance on one or more benchmarks. By tailoring physical relaxation timescales -- and specifically by engineering the correct separation between $T_2$ and $T$ -- photonic neuromorphic processors can be custom-engineered to match the topological demands of specific complex time-series tasks, establishing timescale engineering as a core design paradigm for hardware-adaptive optical computing.

\subsection{Impact of Spontaneous Emission Noise on Reservoir Stability}
\begin{figure}[ht!]
\centering
\includegraphics[width=6.5cm]{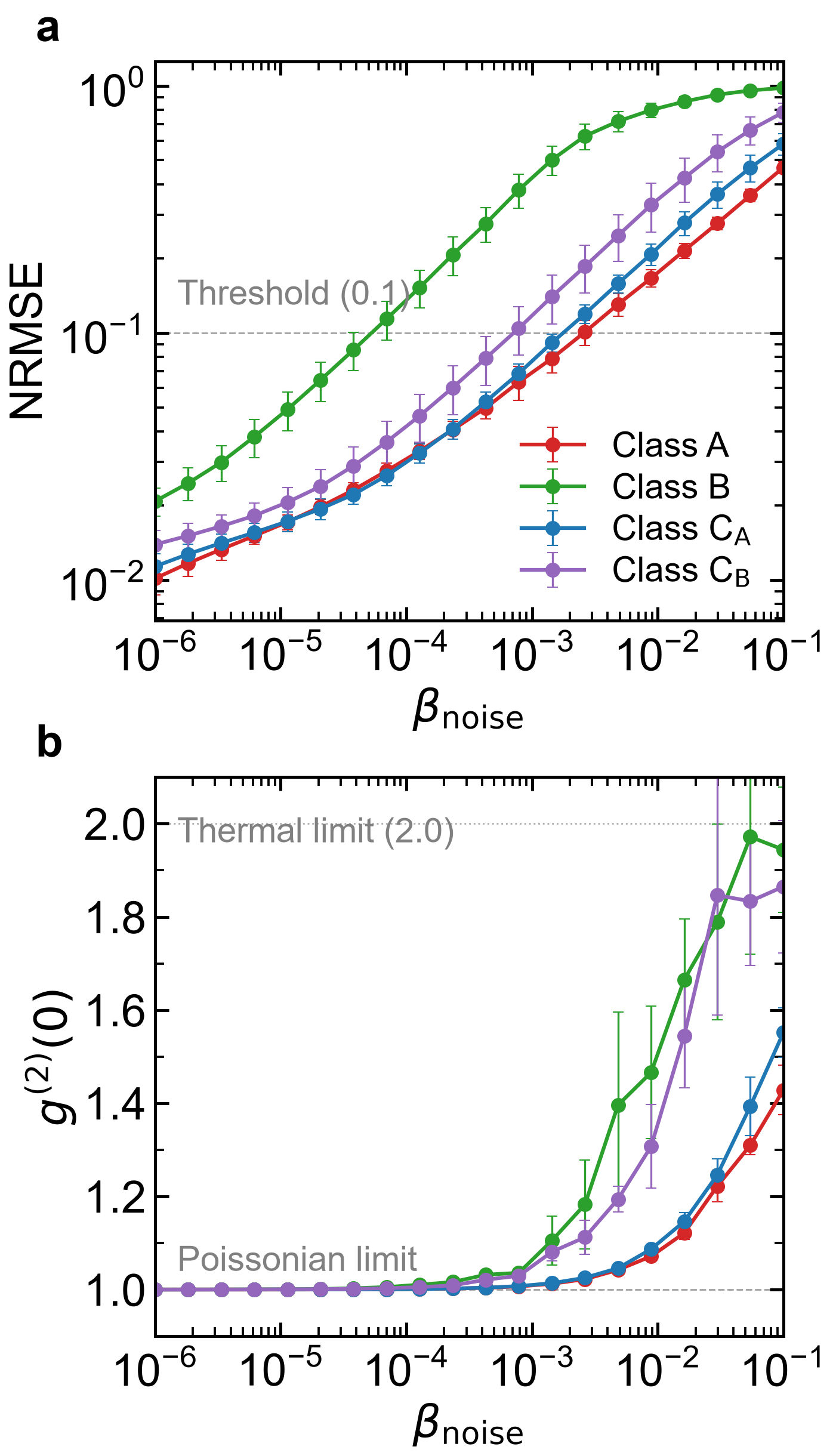}
\caption{Robustness of the optimized TDRC systems against intrinsic spontaneous emission noise. (a) Average NRMSE for the Mackey-Glass one-step prediction task as a function of the spontaneous emission factor $\beta_{\mathrm{noise}}$, spanning $10^{-6}$ to $10^{-1}$ on a logarithmic scale. Each curve corresponds to one laser class: Class A (red), Class B (green), Class C$_A$ (blue), and Class C$_B$ (purple). Error bars indicate the standard deviation across 10 independent mask realizations. The dashed gray horizontal line marks the engineering target threshold of NRMSE = 0.1. (b) Corresponding second-order intensity correlation $g^{(2)}(0)$ of the free-running lasers (without input modulation). The dashed gray line marks the 
Poissonian (coherent) limit $g^{(2)}(0) = 1$; the dotted gray line marks the thermal limit $g^{(2)}(0) = 2$. Class A exhibits the slowest 
degradation of coherence, reaching only $g^{(2)}(0) = 1.43$ at $\beta_{\mathrm{noise}} = 10^{-1}$, whereas the long-carrier classes (Class B and Class C$_B$) approach the thermal limit more rapidly. The finite polarization lifetime $T_2 = 2.0$ in Class C$_B$ provides a modest but measurable protection compared with Class B, as reflected in both panels. Note that the NRMSE values in panel (a) increase smoothly and monotonically for all four classes, without any sharp threshold-like transition.}
\label{Beta_Noise_Robustness_Scan}
\end{figure}

To evaluate physical scalability under realistic quantum noise, we stress the optimized TDRC systems with varying spontaneous emission factors $\beta_{\text{noise}}$ ($10^{-6}$ to $10^{-1}$). Fig.~\ref{Beta_Noise_Robustness_Scan} presents both the prediction accuracy (a) and the intrinsic coherence degradation (b), revealing a critical threshold at $\beta_{\text{noise}} \approx 10^{-3}$. Below the noise threshold, all four laser classes exhibit deterministic behavior. Beyond this value, the intrinsic carrier lifetime $T$ dominates the diverging fates.

Class A (red) displays the highest resilience to quantum noise, with NRMSE rising only to $0.47$ even at the extreme limit $\beta_\mathrm{noise} = 10^{-1}$. The exceptionally fast carrier dynamics ($T = 0.01$) act as a broadband noise sink: spontaneous emission fluctuations are dissipated on a timescale much shorter than the symbol period, so they are naturally filtered out before the next virtual node is sampled. This is directly confirmed by the corresponding intensity correlation, which remains close to the coherent limit ($g^{(2)}(0) = 1.43$ at $\beta_\mathrm{noise} = 10^{-1}$, the lowest among all classes).

Class C$_A$ (blue) exhibits the second-best noise resilience (NRMSE $= 0.58$ at $\beta_\mathrm{noise} = 10^{-1}$), with a smooth monotonic increase as a function of $\beta_\mathrm{noise}$ and no evidence of a threshold-induced "coherence avalanche." The finite polarization lifetime ($T_2 = 2.0$) provides a mild additional filtering effect, but the dominant noise-robustness mechanism remains the finite carrier lifetime ($T = 2.0$), which dissipates fluctuations more quickly than the long-carrier classes.

Class C$_B$ (purple) achieves intermediate noise resilience (NRMSE $= 0.78$ at $\beta_\mathrm{noise} = 10^{-1}$), but a residual $T_2$-mediated benefit is clearly visible. Despite sharing the same ultra-long carrier lifetime as Class B ($T = 392$), Class C$_B$ outperforms Class B by 20\% at the extreme noise limit ($0.78$ vs.\ $0.98$). The finite polarization lifetime ($T_2 = 2.0$) acts as an inline low-pass filter that partially blocks high-frequency spontaneous emission before it can be integrated by the long-lived carriers. This effect is manifested in the intensity correlation, which reaches $1.87$ for Class C$_B$ versus $1.94$ for Class B at $\beta_\mathrm{noise} = 10^{-1}$.

Class B (green) displays the poorest noise resilience (NRMSE $= 0.98$ at $\beta_\mathrm{noise} = 10^{-1}$), acting as an unbuffered noise integrator. The ultra-long carrier lifetime ($T = 392.0$) causes spontaneous emission fluctuations to be integrated over very long timescales, driving the intensity correlation towards the thermal limit ($g^{(2)}(0) \to 2.0$). The absence of any intrinsic filtering mechanism means the noise accumulates monotonically with the pump modulation.

These stress tests establish a fundamental physical trade-off for photonic neuromorphic design. The dominant determinant of noise resilience is the carrier lifetime: short-carrier classes naturally dissipate noise faster than they can be integrated, whereas long-carrier classes suffer from progressive noise accumulation. Against this dominant background, a finite polarization lifetime $T_2$ provides a systematic but secondary (approximately 20\%) enhancement by acting as an inline low-pass filter, which is most valuable precisely when the carrier lifetime is long enough for the polarization filter to be the only available protection mechanism. The design principle that emerges from these results is therefore not simply ``retain a finite $T_2$,'' but rather ``engineer the polarization lifetime in concert with the carrier lifetime to achieve the specific time-scale hierarchy demanded by the target application.''

\section{Conclusion and Discussion}
In conclusion, we have established time-scale engineering as a nuanced design paradigm for photonic reservoir computing. Our unified theoretical framework reveals that retaining a finite polarization lifetime $T_2$ fundamentally reshapes the phase-space topology, information processing capacity, and noise immunity of TDRC systems.

This comprehensive study exposes a fundamental task-dependent reversal that defines the design space of photonic neuromorphic hardware:

(i) Short-horizon nonlinear mapping favors fast carriers. Class A and Class \(C_A\) achieve comparable, and the lowest, NRMSE on the short-horizon Mackey--Glass task (0.0103 and 0.0105, respectively; the difference is within one standard deviation), consistent with their operation in the static-nonlinear-mapping regime. Class A retains the highest linear IPC (4.70) and the smaller mask-to-mask spread.

(ii) Long-horizon chaotic prediction favors matched time scales. Class C$_A$ achieves the best Lorenz and Santa Fe performance ($0.0506$ and $0.4251$), consistent with its high nonlinear IPC ($16.28$) and its balanced time-scale hierarchy ($\theta_\mathrm{opt}/\tau_\mathrm{ro} \approx 0.93$,  corresponding to a clock-cycle-to-delay ratio of $N_v \theta_{\text{opt}}/\tau \approx 0.74$).

(iii) The long-carrier classes exhibit systematically reduced performance across all three tasks. Class B is the poorest performer (MG: 0.0217, Lorenz: 0.1544, Santa Fe: 0.6955), and Class \(C_B\) retains a residual advantage over Class B (MG: 0.0141, Lorenz: 0.0763, Santa Fe: 0.5341) due to polarization-induced low-pass filtering, but neither can compete with Class A or Class \(C_A\). On the Santa Fe dataset, Class A (0.5233) and Class \(C_B\) (0.5341) are statistically indistinguishable. We note that the fixed delay \(\tau = 50\) used throughout this study corresponds to \(\tau/T \approx 0.128\) for the long-carrier classes, i.e., only about one-eighth of their carrier lifetime; this fixed short delay specifically limits the memory available to slow-carrier dynamics. A proportionally longer delay would be required for a fully fair comparison of these classes, which we leave for future work.

(iv) Under realistic quantum noise, the carrier lifetime \(T\) is the dominant determinant of noise resilience. Class A retains the lowest NRMSE (0.47) even at \(\beta_{\rm noise} = 10^{-1}\), whereas Class B degrades to 0.98. A residual \(T_2\)-mediated benefit of approximately \(20\%\) persists in the slow-carrier Class \(C_B\) (0.78 vs. 0.98).

By mapping abstract relaxation dynamics to actionable device-level parameters (e.g., spin-relaxation rates in VCSELs or photon lifetimes in high-$\beta$ microcavities), our findings provide concrete physical design principles for engineering fault-tolerant, high-throughput optical computing chips. We emphasize that the deliberate retention of a finite polarization lifetime is a powerful, practical lever, but that its benefit is maximized only when the polarization lifetime is engineered in the correct relationship to the carrier lifetime and to the task-specific time-scale demands.

\section*{Acknowledgements}
TW thanks the support from National Natural Science Foundation of China (Grant No. 62475206), and Key Research and Development Plan of Shaanxi Province of China (Grant No. 2024GH-ZDXM-42). KL acknowledges support  from the European Union's Horizon 2020 programme under grant agreement number 101129904,
SPIKEPro.



\bibliography{biblio}

\end{document}